\documentclass[aps,prl,twocolumn,showpacs]{revtex4}
\usepackage{graphicx}

\usepackage{ulem}

\begin{document}

\title{All-optical memory based on the injection locking bistability of a two-color laser diode}

\author{S. Osborne, K. Buckley, A. Amann, and S. O'Brien}

\affiliation{Tyndall National Institute, University College, Lee Maltings, Cork, Ireland}

\begin{abstract}
We study the injection locking bistability of a specially engineered two-color semiconductor Fabry-P\'erot 
laser. Oscillation in the uninjected primary mode leads to a bistability of single mode and two-color equilibria. 
With pulsed modulation of the injected power we demonstrate an all-optical memory element based on this 
bistability, where the uninjected primary mode is switched with 35 dB intensity contrast. Using experimental 
and theoretical analysis, we describe the associated bifurcation structure, which is not found in single 
mode systems with optical injection. 
\end{abstract}


\maketitle

\section{Introduction}

Much current research in optical device physics is focused on
bistable, switchable semiconductor lasing elements. The aim is to
provide sophisticated functions for all-optical signal processing,
including optical memory, flip-flop or optical logic [1-4].
The systems considered comprise single \cite{raburn_06} or coupled 
\cite{hill_04, jeong_06} devices including micro-cavity lasers 
\cite{wang_08}, edge-emitting devices \cite{ramos_05} and VCSELs [8-10].
Although laser diodes are known to exhibit complex dynamical features, 
they are nevertheless successfully modeled by surprisingly 
low-dimensional and generic systems of equations. Laser diodes therefore 
provide model systems for the understanding of fundamental concepts of 
nonlinear dynamics, such as non-invasive chaos control \cite{schikora_06} 
or synchronisation of coupled chaotic systems \cite{fischer_00}.

Switching between co-existing states of bistable semiconductor laser
devices can be achieved using modulated or pulsed optical
injection. The structure of the bifurcations in \textit{single mode}
semiconductor lasers with optical injection are well understood. For a
recent review see reference \cite{wieczorek_05}. While two-mode 
models have successfully explained the switching between single 
mode states with different polarizations in VCSELs 
\cite{gatare_06}, a general understanding of nonlinear 
dynamical phenomena in laser diodes for the case of more than one
simultaneously active mode is still lacking. In edge-emitting lasers a
reason for this is that plain edge-emitting Fabry-P\'erot (FP) laser
diodes will generally operate with many longitudinal modes close to
threshold, which complicates the interpretation and modeling of
experimental results considerably \cite{simpson_97, white_98}.

Using an inverse problem solution \cite{obrien_06a}, which relates the threshold 
gain modulation in wavenumber space to the spatially varying refractive index in 
a FP laser, we have designed a two-color FP laser diode with THz primary mode 
spacing. The large engineered mode spacing of this two-color device leads to weakly 
coupled modes with the result that the bias current can be adjusted so
that the two primary modes oscillate simultaneously with the same average power 
level. With only two active modes, such a device represents a test-bed for the 
systematic study of the nonlinear dynamics of multiwavelength semiconductor lasers. 

In this paper, we examine the locking bistability of the two-color device with 
optical injection. In contrast to the case of a single mode laser, where a bistability 
is found between a stable limit cycle associated with wave mixing and the injection 
locked state \cite{wieczorek_05, kovanis_99}, we demonstrate that in the two-color 
laser a bistability between the injection locked (single mode) state and a two-color 
steady state is possible. We experimentally demonstrate all-optical switching 
between these bistable states, where the uninjected mode of the device performs 
an inverted memory function with greater than 35 dB intensity contrast. We explain 
the underlying mechanism using modeling and bifurcation analysis, showing that a 
four-dimensional model of the system provides excellent agreement with experimental 
results. Our results demonstrate a new and unique approach to in-plane all-optical 
signal processing. 

\section{Experimental results}

The device we consider is a multi-quantum well $\mbox{InP}/\mbox{In}_{x}\mbox{Ga}_{y}\mbox{Al}_{1-x-y}\mbox{As}$ 
FP laser of length 350 $\mu$m with a peak emission near 1.3$\mu$m. It incorporates slotted regions 
etched into the laser ridge waveguide that determine the lasing mode spectrum. Further details of the design 
and free running lasing characteristics of the device can be found in \cite{obrien_06b}. 
Adjusting the device current, we equalize the time averaged optical power in each primary mode of the free 
running laser. The primary mode spacing between the short wavelength mode $\nu_{1}$ and the long wavelength 
mode $\nu_{2}$ is chosen as four fundamental FP modes (480 GHz).

\begin{figure}
\includegraphics[width=0.9\columnwidth]{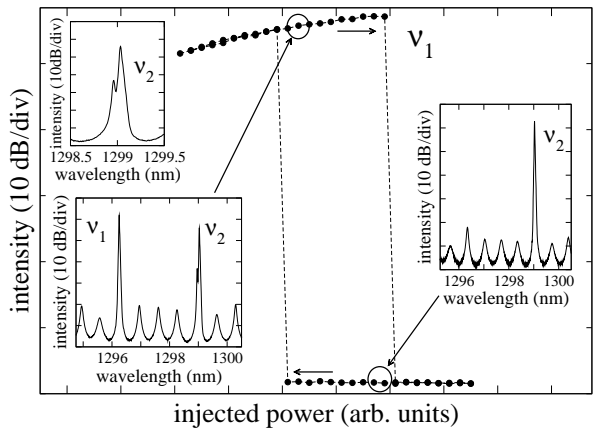}
\caption{\label{opex1} The optical power of the short wavelength mode, $\nu_{1}$, as a function of the 
injected power at the long wavelength mode, $\nu_{2}$, in up and down sweep as indicated by the 
arrows. The lower left inset shows the two color equilibrium state. The upper left inset shows the injected 
mode in this state in close-up where a two-peak structure is seen. These peaks are separated by the 
detuning frequency. The right inset shows the single mode locked state of the device.} 
\end{figure}

For our experiment, we use the two-colour laser in a master-slave configuration, where the master laser is a 
tunable laser with $< 100$ kHz linewidth. A hysteresis loop of the light output power in mode $\nu_{1}$ for 
increasing and decreasing injected power in mode $\nu_{2}$ is shown in Fig. \ref{opex1}. In the insets of Fig. 
\ref{opex1} we show the co-existing states associated with the locking bistability. The injected mode 
$\nu_{2}$ is at long wavelength and the detuning in this case is $\Delta \omega = -14$ GHz, which is 
approximately 3.5 times the relaxation oscillation frequency, $\nu_{\mbox{\tiny RO}} = 3.9$ GHz. For weak 
injection the dynamics are two mode and time-periodic due to the wave-mixing response. At large  injection, 
single mode oscillation with locking of the output to the injected light is observed. 
Decreasing the optical injection from inside the locking region, the single mode locking state and a 
two-color state are found to co-exist. In the two-color state we find that the uninjected 
mode has a greater intensity than the injected mode, whose spectrum (upper left inset of Fig. \ref{opex1}) 
comprises two distinct peaks, which are separated by the detuning of the injected light frequency from 
the free running cavity mode frequency.  

\begin{figure}
\includegraphics[width=0.9\columnwidth]{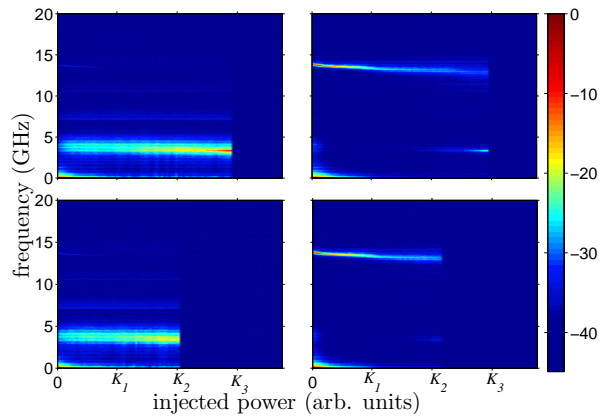}
\caption{\label{opex2}  Power spectral densities (PSD) of each of the primary modes of the device 
as a function of the injected power. The upper panel shows the PSD for increasing injection, while the 
lower panels show the PSD for decreasing injection. The frequency detuning is $\Delta \omega =$ -14 GHz and the 
relaxation oscillation frequency is $\nu_{\mbox{\tiny RO}} =$ 3.9 GHz. 
Left panels: uninjected mode, $\nu_{1}$. Right panels: Injected mode, $\nu_{2}$.}
\end{figure}

The optical spectra in Fig. \ref{opex1} are complemented by the mode resolved power spectral densities
depicted in Fig. \ref{opex2}. We observe four regions of qualitatively different dynamics, the boundaries 
between which we have indicated using corresponding values of injected field strength $K_{i}, \;i = 1, 2, 3$.  
At weak injected power (injected field strength $K < K_{1}$) we find a two-mode wave-mixing region 
where the detuning frequency is observed in the dynamics of the injected mode and a weak feature at 
the relaxation oscillation frequency is observed in the dynamics of the uninjected mode. 
With increasing injected power ($K_{1} < K < K_{3}$) this feature in the uninjected mode 
becomes considerably stronger and at the same time the peak at the detuning frequency 
in the dynamics of the injected mode broadens. At large injected power ($K> K_{3}$) we enter the single
mode locking region where the spectrum is structureless. For $K_{2} < K < K_{3}$ we identify a region
of optical bistability between a two-mode and a single mode state. The selected stable state depends on 
the sweeping direction.

\begin{figure}
\includegraphics[width=0.9\columnwidth]{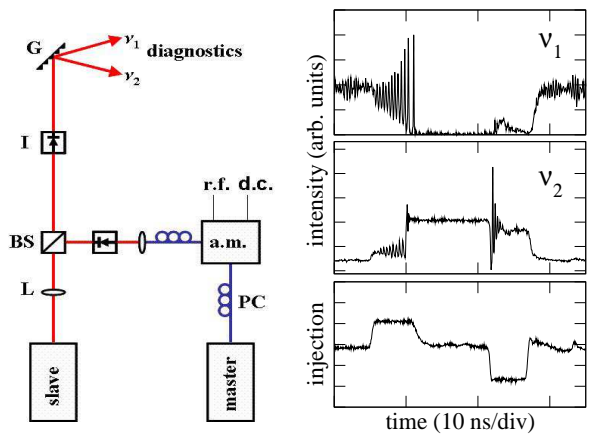}
\caption{\label{opex3} Left panel: The experimental setup schematic for the all-optical memory 
element device operation; PC, polarization controller; A.M., amplitude modulator;  BS, beam splitter; L, lens;
I, isolator; G, grating. Right panels: The intensity time traces of the uninjected mode, $\nu_{1}$, the 
injected mode, $\nu_{2}$, and of the injected field strength. The pulse duration is 5 ns.}
\end{figure}

We have been able to switch between these two bistable states using an appropriate pulsed modulation of the 
injected signal. The experimental setup and the mode resolved time traces in response to optical injection 
modulation with 5 ns pulse width are shown in Fig. \ref{opex3}. One can see that the device operates as a memory 
element where we can regard the mode at short wavelength as performing the (inverted) memory function and the 
injected mode at long wavelength as the writing channel. In particular, the intensity of the uninjected mode 
is switched between an ``off'' state and an ``on'' lasing state  with a contrast ratio of greater than 35 dB. 

\section{Modeling and bifurcation analysis}

To understand the structure of the bifurcations that lead to this bistability we have adapted the 
well known model of a single mode laser \cite{heil_01} with optical injection 
to account for the presence of a second lasing mode. The system of equations in normalized units 
may be written as follows:
\begin{eqnarray} 
&\dot{E}_{1} =  \frac{1}{2} (1 + i\alpha)(g_{1}(2n + 1) - 1)E_{1} \label{eq:1}\\
&\dot{E}_{2} =  \left[\frac{1}{2} (1 + i\alpha)(g_{2}(2n + 1) - 1) - i\Delta\omega\right]E_{2} + K  \label{eq:2}\\
&T\dot{n} = P - n - (1 + 2n)\sum_{m}g_{m}|E_{m}|^{2} \label{eq:3} 
\end{eqnarray}
where the nonlinear modal gain is 
\begin{equation}
g_{m} = g_{m}^{(0)}\left(1 + \epsilon\sum_{n}\beta_{mn}|E_{n}|^{2}\right)^{-1}.
\end{equation}
In the above equations, $g_{m}^{(0)}$ is the linear modal gain while the $\epsilon\beta_{mn}$ determine 
the cross and self saturation. The normalized complex electric field amplitudes of the modes are given by 
$E_{m}$ and the normalized excess carrier density is $n$. $T = \gamma \tau_{s}$, where $\tau_{s}$ is the  
carrier lifetime and  $\gamma$ is the cavity decay rate. The phase-amplitude coupling is given by $\alpha$ while
$P$ is the normalized pump current. The bifurcation parameters are the normalized injected field strength 
$K$ and the angular frequency detuning $\Delta\omega$.

Parameters used are $\alpha = 2.6$, $P = 0.25$, (50 \% above threshold), $T^{-1} = 0.00125$, $\epsilon = 0.01$, 
$\beta_{12} = \beta_{21} = 2/3$ and $\beta_{11} = \beta_{22} = 1$. This ratio between the cross and self 
saturation is consistent with the stability of the two-mode solution in the free running laser
\cite{obrien_06b}. 

Note that although we have provided a complex equation for the field
$E_{1}$, the phase of $E_{1}$ is in fact decoupled leading to a
four-dimensional system of equations. Only the intensity of the
uninjected mode influences the dynamics, reflecting the fact that
the mode spacing is in the highly non-degenerate regime. More generally, 
our model describes two oscillators, whose amplitudes
are coupled via a global variable $n$, and where one of the 
oscillators is forced by a periodic external input.

\begin{figure}
\includegraphics[width=0.9\columnwidth]{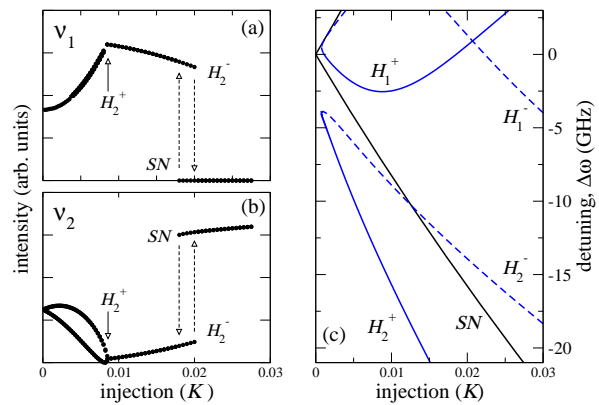}
\caption{\label{opex4} Local extrema of the field intensities (a) $|E_1|^2$ and (b) $|E_2|^2$ 
obtained from numerical integration of Eqs. (\ref{eq:1})--(\ref{eq:3}) as a function of the injected field 
strength for increasing and decreasing injection. The frequency detuning is -14 GHz. (c) Bifurcation diagram 
in the $\Delta \omega$ vs. $K$ plane. Subcritical and supercritical Hopf bifurcations are indicated by dashed 
and solid blue lines and labeled via $H_m^-$ and $H_m^+$, respectively. $H_1^\pm$ and $H_2^\pm$ denote Hopf
bifurcations of single and two mode equilibria, respectively. The solid black line $SN$ is the saddle-node
bifurcation of the single mode locked state.}
\end{figure}

Using the model equations (\ref{eq:1})--(\ref{eq:3}), we can uncover the
bifurcation scenario which is responsible for the observed bistable behaviour. In Fig. \ref{opex4} (a) and 
(b) the local extrema of the numerically calculated field intensities $|E_1|^2$ and $|E_2|^2$ are
shown for increasing and decreasing $K$ and fixed $\Delta \omega$. The bifurcations evident in Fig. 
\ref{opex4} (a) and (b) can be understood from the global bifurcation diagram in the $\Delta \omega$ vs. 
$K$ plane shown in Fig. \ref{opex4} (c), which was calculated using the numerical continuation tool AUTO-07p 
\cite{auto}. As $K$ increases from zero, the stable two mode limit cycle associated with $\Delta
\omega$ which exists for small $K$ evolves into a stable two mode equilibrium via a supercritical Hopf 
bifurcation at $H^+_2$. With increasing $K$ this two mode equilibrium becomes unstable via the
subcritical Hopf bifurcation at $H_2^-$. On the other hand, however, the single mode locked state becomes 
stable via a saddle-node bifurcation at the line $SN$ (cf. Fig. \ref{opex4} (c)). Bistability therefore 
occurs if the $SN$ bifurcation occurs at a smaller $K$ value than the $H^-_2$ bifurcation, which is the 
case for sufficiently negative detuning ($\Delta \omega < -10$GHz) as evident from Fig.\ref{opex4} (c). 
This explains the hysteresis loops found in Fig. \ref{opex4} (a) and (b).

We are now in a position to connect our numerical findings in Fig. \ref{opex4} 
with the experimental results of Fig. \ref{opex2}. For sufficiently large $\Delta \omega$, 
the four eigenvalues of the Jacobian matrix of the two-mode equilibrium comprise two 
complex conjugate pairs with imaginary parts equal to the detuning frequency $\Delta \omega$ 
and the relaxation oscillation frequency $\nu_{\mbox{\tiny RO}}$. 
We identify the boundary $K_1$ in Fig. \ref{opex2} with the supercritical Hopf bifurcation
$H_2^+$ in Fig. \ref{opex4}. At $H_2^+$ the real parts of the eigenvalues associated 
with $\Delta \omega$ change sign and become negative with increasing $K$. Experimentally this 
is reflected in a significant broadening of the spectral feature at frequency $\Delta \omega$ 
in the dynamics of the injected mode. This is consistent 
with the expected behaviour in a noisy system and explains the observed double peak 
structure in the optical spectrum shown in the upper left inset of Fig. \ref{opex1}. 
We identify the upper boundary of the bistable region $K_3$ in Fig. \ref{opex2} with 
the subcritical Hopf bifurcation $H_2^-$ of Fig. \ref{opex4}. At $H_2^-$ the real parts 
of the eigenvalues associated with $\nu_{\mbox{\tiny RO}}$ change sign. Within the bistable 
region, spontaneous emission noise is able to excite precursor oscillations at 
$\nu_{\mbox{\tiny RO}}$ which are visible in the power spectra of both modes in Fig. \ref{opex2}. 
Finally, the lower end of the bistable region $K_2$ in Fig. \ref{opex2} can be identified with 
the saddle-node bifurcation $SN$ in Fig. \ref{opex4}. When approaching this bifurcation with
decreasing $K$, no precursor oscillations are observed, as  saddle-node bifurcations do not 
generically involve oscillatory behaviour. All of the above conclusions are confirmed by 
numerical simulations that include spontaneous emission noise. 

It is interesting to compare this bifurcation structure with others
considered in the literature. In \cite{wieczorek_05} a bistability
between a limit cycle and a locked state was reported in optically
injected single mode lasers. This limit cycle becomes unstable via a
subcritical torus bifurcation. In contrast, our bistability mechanism
involves only Hopf and saddle-node bifurcations and occurs between two
equilibria. The relative simplicity of our bifurcation scenario is
therefore a result of the presence of the second lasing mode, and the
increase of the dimensionality of the dynamical system by one. A pair 
of subcritical and supercritical Hopf bifurcations have also been 
associated with relaxation-oscillation and wave-mixing pulsations in 
the context of in-plane semiconductor lasers with integrated optical 
feedback \cite{ushakov_04} where similar precursor features close to 
the bifurcation points were also reported.

The bifurcations of a two-mode system have also been studied in the
context of optically injected VCSELs to explain the switching between
single-mode states with different polarizations \cite{gatare_06}. 
However, since the two polarization modes in a VCSEL have a very small
frequency spacing (c.\ 10 GHz), these polarization modes are intrinsically 
bistable, and a six dimensional system of equations
is required for adequate modeling. As a result the bifurcation
scenario differs significantly from that considered here, which involves 
an equilibrium state with two simultaneously active and highly nondegenerate 
lasing modes. We note also that the case of two-mode injection locking in an 
optically injected VCSEL \cite{sciamanna_05} must also be contrasted with our 
case, as injection locking implies a frequency degeneracy between the 
orthogonally polarized modes and the injected field. Such a state, which 
can be regarded as  a single mode state with elliptical polarization, cannot 
arise in the context of our model as the phase of the uninjected mode is 
decoupled. 

\begin{figure}
\includegraphics[width=0.9\columnwidth]{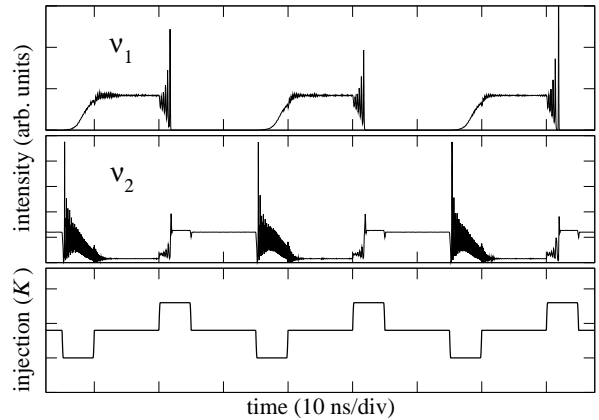}
\caption{\label{opex5} Numerical intensity time traces of the injected field strength and of the two 
primary modes of the device. The detuning is -14 GHz. The average value of the injected field strength 
is 0.018 with a modulation of $\pm$ 0.008. The pulses are 5 ns duration with a 200 ps rise-time.}
\end{figure}

In Fig. \ref{opex5} we show the results of numerical modeling of the optical memory element. 
For this simulation we have used a 5 ns pulse duration and we also included spontaneous emission 
noise. Agreement with the measured intensity time traces is very good. In particular, we have 
reproduced the large and decaying relaxation oscillations in the injected mode's unlocking 
dynamics. The fast locking of the injected mode and the fast extinction of the uninjected 
mode is also reproduced as is the rather slower recovery of the uninjected mode once the 
injected mode unlocks. Numerical results suggest that the memory element can be switched 
between states with subnanosecond pulses. 

\section{Conclusions}

We have demonstrated an all-optical memory element based on the injection locking of a 
two-color Fabry-P\'erot laser diode. Pulsed modulation of the injected field leads to switching of the 
uninjected mode with 35 dB intensity contrast. The underlying bistability in the system was found to be 
between the single mode injection locked state and a two-color equilibrium state, unlike in the single 
mode injected laser. Numerical modeling of the switching dynamics was in excellent agreement with
experimental results. The simplicity of our approach should lead to new opportunities 
for all-optical signal processing with in-plane semiconductor lasers.  

\section*{Acknowledgements}

This work was supported by Science Foundation Ireland and IRCSET. The authors thank G. Huyet and
S. P. Hegarty for helpful discussions and Eblana Photonics for the preparation of sample devices.

\end{document}